\documentclass[10pt,aps,pre,amsmath,amsfonts,amssymb,floatfix,longbibliography,superscriptaddress,notitlepage, twocolumn]{revtex4-1}
\usepackage{mathrsfs} \usepackage{graphicx,color} \usepackage{bbm} \usepackage{multirow}
\usepackage[normalem]{ulem}
\usepackage{lineno}
\usepackage{empheq}

  \def\erf{\text{erf}}

\def\to{\rightarrow}

\newcommand{\beq}{\begin{equation}} \newcommand{\eeq}{\end{equation}}

\renewcommand{\phi}{\varphi}

\renewcommand{\vec}[1]{\boldsymbol{\mathrm{#1}}}

\newcommand\YD[2]{\textcolor{blue}{#2}}

\begin{document}
\title{Jamming is a first-order transition with quenched disorder 
in amorphous materials sheared by cyclic quasistatic deformations}

\author{Yue Deng}
\affiliation{
Institute of Theoretical Physics, Chinese Academy of Sciences, Beijing 100190, China}
\affiliation{School of Physical Sciences, University of Chinese Academy of Sciences, Beijing 100049, China}

\author{Deng Pan}
\affiliation{CAS Key Laboratory of Theoretical Physics, Institute of Theoretical Physics, Chinese Academy of Sciences, Beijing 100190, China}

\author{Yuliang Jin}
\email{yuliangjin@mail.itp.ac.cn}
\affiliation{CAS Key Laboratory of Theoretical Physics, Institute of Theoretical Physics, Chinese Academy of Sciences, Beijing 100190, China}
\affiliation{School of Physical Sciences, University of Chinese Academy of Sciences, Beijing 100049, China}
\affiliation{Center for Theoretical Interdisciplinary Sciences, Wenzhou Institute, University of Chinese Academy of Sciences, Wenzhou, Zhejiang 325001, China}

\begin{abstract}
Jamming is an athermal transition between flowing and rigid states in amorphous systems such as granular matter, colloidal suspensions, complex fluids and cells. 
The jamming transition seems to display mixed aspects of a first-order transition, evidenced by a discontinuity in the coordination number, and a second-order transition, indicated by power-law scalings and diverging lengths. Here we demonstrate that jamming is a first-order transition with quenched disorder in cyclically sheared systems with quasistatic deformations{, in two and three dimensions}. 
{Based on scaling analyses, we show that}
fluctuations of the jamming density in finite-sized systems have important consequences on the finite-size effects of various quantities,  resulting in a square relationship between disconnected and connected susceptibilities, a key signature of the first-order transition with quenched disorder. This study puts the jamming transition into the category of a broad class of transitions in disordered systems {where sample-to-sample fluctuations dominate over thermal fluctuations}, {suggesting that the nature and behavior of the jamming transition 
{might be better understood}
within the developed theoretical framework of the athermally driven random-field Ising model.}


\end{abstract}

\maketitle

{\Large \bf Introduction}

Jamming in athermal particles is a paradigm of transitions between fluids and amorphous solids~\cite{makse2000packing, o2003jamming, liu2010jamming, van2009jamming, PAN20231}, with a deep connection to the glass transition in thermal systems~\cite{parisi2010mean, charbonneau2017glass, parisi2020theory}.  
Recent studies have revealed 
extremely rich features in the jamming phenomenon, but the   nature of the jamming transition remains inconclusive.

(i) {\it Mechanical marginality, related diverging length scales and power-law scalings.}
At the jamming transition density (volume fraction) $
\varphi_{\rm J}$, the isostatic condition needs to be satisfied for the coordination number $Z$ (average number of contacts per particle): $Z = Z_{\rm iso} = 2d$ in frictionless, infinite systems,   
where $d$ is the dimensionality~\cite{maxwell1870reciprocal, alexander1998amorphous}.
Isostaticity implies that at $\varphi_{\rm J}$ the system  is marginally stable, inspiring the search for diverging length scales. 
According to the ``cutting argument", the isostaticity gives rise to a  diverging { isostatic length} scale at jamming, $l^*\sim \Delta Z^{-1} \sim (\varphi - \varphi_{\rm J})^{-1/2}$, below which mechanical stability of the bulk system is affected by the cutting boundaries~\cite{wyart2005rigidity, wyart2005geometric, silbert2005vibrations}, where $\Delta Z =  Z - Z_{\rm iso}$ is the excess coordination number.
The effective medium theory gives 
a  { scattering length} scale diverging at $\varphi_{\rm J}$, $l_{\rm c} \sim \Delta Z^{-1/2} \sim (\varphi - \varphi_{\rm J})^{-1/4}$, below which continuum elasticity breaks down~\cite{wyart2010scaling, degiuli2014effects, lerner2014breakdown}. 
Other related diverging length scales include the transverse wavelength, $\xi_{\rm T} \sim (\varphi - \varphi_{\rm J})^{-\nu_{\rm T}}$ with $\nu_{\rm T} \approx 0.24$~\cite{silbert2005vibrations}, and the longitudinal wavelength, $\xi_{\rm L} \sim (\varphi - \varphi_{\rm J})^{-\nu_{\rm L}}$ with $\nu_{\rm L} \approx 0.48$~\cite{silbert2005vibrations}.


 At jamming, the marginal stability analysis  provides relationships between the exponents $\theta$ appearing in the power-law distribution of  weak inter-particle forces $P(f) \sim f^\theta$ and $\alpha$ in the distribution of small inter-particle gaps $P(h) \sim h^{-\alpha}$: $\theta = 1/\alpha -2$ for extensive modes~\cite{wyart2012marginal} and $\theta = 1- 2\alpha$ for localized buckling modes~\cite{lerner2013low, degiuli2014force}. 
Above jamming, the marginal boundary between unstable and stable phases is defined by a scaling relation, $\Delta Z \sim (\varphi - \varphi_{\rm J}) ^{1/2}$~\cite{wyart2005rigidity, muller2015marginal}. 
Other scalings have been  established for over-jammed systems near $\varphi_{\rm J}$~\cite{makse2000packing,  o2003jamming, goodrich2016scaling, liu2010jamming, baule2018edwards, PAN20231}.
For example, a relationship between the shear modulus $G$ and $\Delta Z$ can be derived by microscopic elastic theories, $G \sim \Delta Z \sim  (\varphi - \varphi_{\rm J}) ^{1/2}$~\cite{wyart2005rigidity, zaccone2011approximate,  pan2023shear}.

(ii) {\it Hyperuniformity and associated diverging length scales.} Recent studies reveal that the spatial distribution of the single-particle contact number, $Z_i$, is hyperuniform at jamming.
{The contact hyperuniformity is established  by two power-law scalings measured in simulated packings~\cite{hexner2018two, hexner2019can}: the scaling of the fluctuations of the average contact number in a hyper-cube of volume $\ell^d$, $\sigma_Z(\ell) \sim \ell ^{-\mu}$ with $\mu=1$, and the small wave-vector scaling of the contact number structure factor, $S_Z(q) \sim q^{\alpha_Z}$ with $\alpha_Z \approx 1.53$. The crossover from the hyperuniform regime ($\ell \ll \xi_{\rm H}$) to the uniform regime ($\ell \gg \xi_{\rm H}$) defines a  { hyperuniform correlation length} $\xi_{\rm H}$, which diverges at the jamming transition,} 
{$\xi_{\rm H} \sim \Delta Z^{-\nu_{\rm H}}$}. The value of the exponent {$\nu_{\rm H}$} appears to depend on how to extract the correlation length and the dimensionality.

(iii) {\it Gardner glass phase, landscape marginality and associated criticality.} 
The marginality at jamming has been established by an independent approach within 
the framework of replica symmetry breaking~\cite{charbonneau2014fractal, parisi2020theory, charbonneau2015numerical, berthier2016growing,  berthier2019gardner, li2021determining, urbani2023gardner}.
Unjammed hard sphere glasses undergo a { Gardner transition} where the free energy landscape becomes fractal and marginal, and the caging susceptibility diverges~\cite{berthier2016growing}. The entire Gardner phase, including the jamming limit, is critical.   
In other words, the { caging correlation length},
{$\xi_{\rm G} \sim \infty$},
remains infinite near jamming, {when the transition is approached ($\varphi \to \varphi_{\rm J}$) from below ($\varphi < \varphi_{\rm J}$)}.
The mean-field replica theory predicts the values of exponents in the weak force and small gap distributions, 
{$\alpha  = 0.41 \ldots$}
and 
{$\theta  = 0.42 \ldots$},
coinciding with the relationship $\theta = 1/\alpha -2$ given by the mechanical marginal stability analysis~\cite{charbonneau2014fractal, charbonneau2015jamming, wang2022experimental}. The theory provides an additional scaling relationship between the cage size $\Delta$ and the entropic pressure $p$, $\Delta \sim p^{-\kappa}$ with
{$\kappa  = 1.41 \ldots$}
The exponents appear to be 
independent of the dimensionality $d$ for $d \geq d_{\rm u}${~\cite{charbonneau2014fractal, charbonneau2015jamming, parisi2018robustness}}, where $d_{\rm u}=2$ is the conjectured upper critical dimension {of the jamming transition}~\cite{goodrich2012finite}.

(iv) {\it Criticality in shear rheology.}
The criticality of the jamming transition is suggested by scaling analyses of the rheological data obtained in finite-rate shear simulations of flowing states near $\varphi_{\rm J}$~\cite{olsson2007critical, olsson2011critical, kawasaki2015diverging}. Combing the power-law divergence of the viscosity, $\eta = (\varphi_{\rm J} -\varphi)^{-\beta}$ and the vanishing of yield stress, $\sigma_{\rm Y} \sim (\varphi -\varphi_{\rm J})^{\Delta}$, a critical scaling function is proposed. A diverging { rheological correlation length} can be extracted from velocity correlations~\cite{olsson2007critical} or non-affine displacements~\cite{heussinger2009jamming}, $\xi_{\rm R} \sim (\varphi_{\rm J} - \varphi)^{-\nu}$, with {$\nu \approx 1$
~\cite{olsson2007critical, vaagberg2011finite, olsson2020dynamic}}.
The criticality seems to be at odds with the hyperuniformity discussed in (ii) -- in the thermodynamical limit, the fluctuations diverge in the former and vanish in the latter.

(v)  
{\it Discontinuity in the coordination number.}  
The coordination number $Z$, which is considered as an order parameter of the jamming transition{~\cite{liu2010jamming}},  jumps discontinuously from $Z=0$ below jamming,  to $Z \geq Z_{\rm iso}$ above, {under quasi-static compression/decompression~\cite{makse2000packing, o2003jamming} or shear~\cite{vinutha2020timescale}}.
{In unjammed states, $Z = 0$ because particles can push each other apart, leaving no overlapping between them. In jammed states, the contact network spans the entire system and the minimum condition to have such networks is $Z=Z_{\rm iso}$. Thus the discontinuous jump in $Z$ at jamming is essentially related to isostaticity.}
Apparently, as the signature of a first-order transition, this discontinuity is inconsistent with the viewpoint of a continuous transition described above. 
{Note that not all physical quantities exhibit a discontinuity at jamming: some quantities, such as the pressure $P$ and the shear modulus $G$, vanish continuously when the jamming transition is approached from above~\cite{makse2000packing, o2003jamming} {(for example, see Ref.~\cite{kumar2016memory}) for the relation between $P$ and $\varphi$)}. 
On the other hand, other quantities, such as the bulk modulus $B$ and the fraction of non-rattlers $f_{\rm NR}$, jump abruptly at jamming, similar to $Z$.
}

It is clear that  various diverging length scales have been suggested throughout the literature. However, none of the lengths in (i-iv) can explain the finite-size scaling behavior of the jamming fraction $F_{\rm J}(\varphi, N)$~\cite{o2003jamming, bertrand2016protocol, baity2017emergent, jin2021jamming}:
the data can be reasonably collapsed by a {scaling form}, $F_{\rm J}(\varphi, N) = \mathcal{F} [ (\varphi - \varphi_{\rm J}) N^{1/2} ]$, valid for both compression and shear jamming in two (2D) and three (3D) dimensions.
In this study, we show that this scaling can be fully explained by a first-order transition scenario of the jamming transition with quenched disorder. The form $(\varphi - \varphi_{\rm J}) N^{1/2}$  originates from the disorder-induced fluctuation of the jamming density itself in finite-sized systems of $N$ particles, which  follows the standard central limit theorem. 
Thus this finite-size scaling  is independent of  isostaticity, marginality, criticality and hyperuniformity.


Three important differences between previous approaches (i-v) 
and ours shall be denoted. 
First, in (i-iii), the jamming limit is approached from one side only, i.e., the over-jammed side ($\varphi > \varphi_{\rm J}$) in (i) and (ii), and the unjammed side ($\varphi < \varphi_{\rm J}$) in (iii).
Here we consider a well-defined ensemble including both over-jammed and unjammed states, whose ratio is essential in the scaling analysis. Note that once an ensemble average is taken, the (v) discontinuity in $Z$  turns into a smooth function, and thus a finite-size analysis becomes essential to see the asymptotic behavior in the thermodynamic limit.  

Second, in conventional compression 
protocols, the generated ensemble  depends  on the initial conditions ~\cite{chaudhuri2010jamming, ozawa2012jamming, jin2021jamming} and the basins of attraction~\cite{bertrand2016protocol, ashwin2012calculations, martiniani2017numerical}. 
Here we instead consider an ensemble prepared by cyclic shear, where the states are sampled by well-controlled dynamics similar to those in thermal systems.  
Recently, the response of amorphous assemblies of particles to cyclic shear has attracted great interest, due to the presence of  a nonequilibrium  phase transition, called the reversible-irreversible (RI) transition~\cite{pine2005chaos, corte2008random, schreck2013particle, nagasawa2019classification, das2020unified}.
Interestingly, the jamming transition 
 lies in the irreversible regime where particle trajectories are asymptotically diffusive~\cite{nagasawa2019classification, das2020unified}. In this study, we restrict our ensemble within the irreversible phase.
 {The jamming density of this ensemble is 
 equivalent to the minimum jamming density, or the { jamming-point} (J-point) density that is obtained by rapid quench~\cite{o2003jamming}.
 {The protocol dependence of the jamming density is not observed in our simulations: the preparation history of the initial condition becomes irrelevant after the first shear cycle,  because a steady state is reached before the strain reversal.}
 }


  
 Third, in our ensemble, we carefully exclude partially crystallized and fragile states with $Z<Z_{\rm iso}$ that are sensitive to mechanical perturbations or protocol parameters, and regard them as unjammed states with $Z=0$. 
 {Any transient and intermediate states generated during shear, which appear before the steady states, are excluded in our analysis.}
 If such states are included, the discontinuity in $Z$ diminishes and the jamming transition looks continuous, similar  to the results reported in Ref.~\cite{heussinger2009jamming} obtained by uniform shear.
Thus we expect the impactibility between the criticality viewpoint in (iv) and our first-order picture originating from finite-rate effects.  
For finite-time scales, the existence of transient states with $Z<Z_{\rm iso}$ can lead to a continuous jamming transition~\cite{olsson2007critical, olsson2011critical, heussinger2009jamming}. However, our results suggest that, after a sufficiently long time, any configurations with $Z<Z_{\rm iso}$ would eventually relax to unjammed states with zero energy and inter-particle contacts. 
{Note that the timescale for the system to attain force balance diverges at the shear jamming transition~\cite{vinutha2020timescale}, which means that the real quasistatic  limit would correspond to extremely small shear rates in large systems.}
\\

\begin{figure*}[!htbp]
  \centering
  \includegraphics[width=0.6\linewidth]{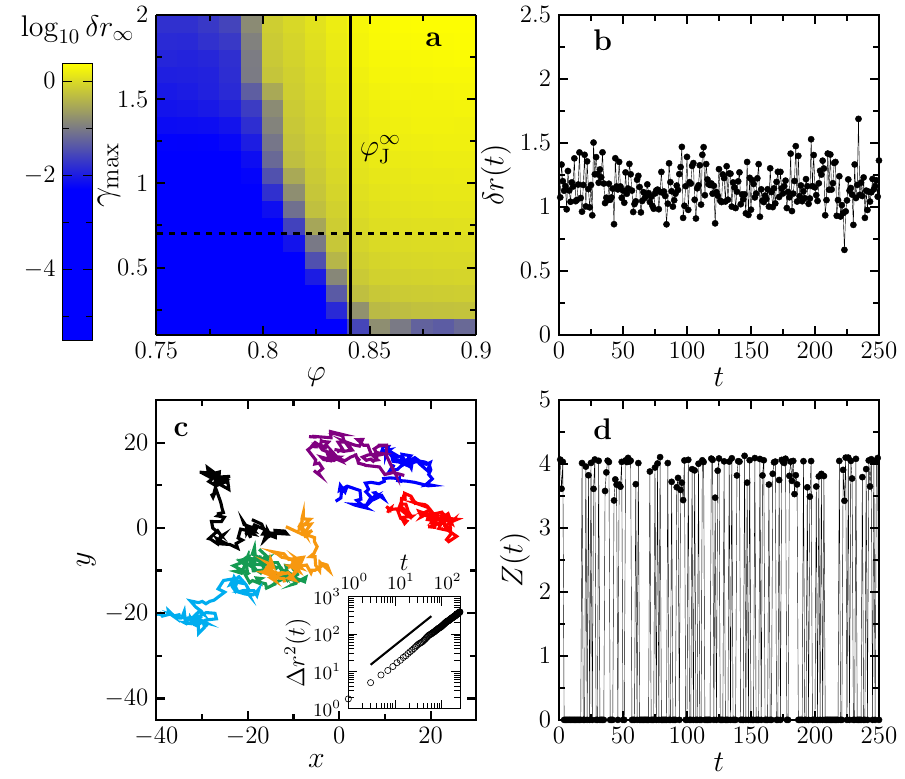}
  \caption{{\bf Cyclic athermal quasi-static shear simulations.} (a) Reversible ($\delta r_{\infty} < 0.1$, blue region) and irreversible ($\delta r_{\infty} > 0.1$, yellow region) phases ($N={1000}$). The solid line represents $\varphi_{\rm J}^\infty = {0.8432}$, and the dashed line represents $\gamma_{\rm max}=0.7$ used to generate the ensemble in the main text. 
  (b) One-cycle displacement $\delta r(t)$ and (d) coordination number $Z(t)$ for a typical sample, {at $\varphi = 0.841$. (c) Several typical particle trajectories during 100 cycles at $\varphi = 0.8425$. Positions are recorded at $\gamma = 0$ only, and the system size is $43.8 \times 43.8$ with periodic boundary conditions. (inset)}
 The MSD data  show diffusive behavior $\langle \Delta r^2(t) \rangle \sim t $ (line).}
  \label{fig:cAQS}
\end{figure*}

{\Large \bf Results}

{\bf An ensemble generated by cyclic athermal quasi-static shear.}

We apply cyclic  athermal quasistatic shear (CAQS) to standard models of soft, frictionless particles in 2D and 3D (see {\it Models} and  {\it Shear protocol} in Methods). We present 2D data in the main text and 3D data in the Supplementary Information (SI).
For the 2D model, previous studies report a jamming density (J-point density) $\varphi_{\rm J} \approx 0.842-0.843$~\cite{o2002random, vaagberg2011finite}.
The phase diagram of RI transitions near 
$\varphi_{\rm J}$ is plotted in Fig.~\ref{fig:cAQS}a.
The RI transition is characterized by the one-cycle displacement averaged over particles~\cite{nagasawa2019classification},
\beq
\delta r(t) = \frac{1}{N} \sum_{i=1}^N \left| \vec{r}_i(t+1) - \vec{r}_i(t) \right|,
\eeq
and the mean-squared displacement (MSD),
\beq
\langle \Delta r^2(t) \rangle = \frac{1}{N} 
\left \langle \sum_{i=1}^N \left| \vec{r}_i(t) - \vec{r}_i(0) \right|^2 \right \rangle,
\eeq
where $t$ is the number of cycles playing a similar role as the time in thermal systems, $\vec{r}_i(t)$ the position of particle $i$ at time $t$ and zero strain $\gamma =0$,
and  $\langle x \rangle$  the average over $N_{\rm s}$ samples.

The RI dynamics near $\varphi_{\rm J}$ are systematically studied in  Ref.~\cite{nagasawa2019classification}, according to which $\delta r(t)$ displays  two-step relaxation behavior typically appearing in  glassy systems. 
For $\tau_{\rm R} < t < \tau_{\rm L}$, $\delta r(t)$
develops a plateau at $\delta r_{\rm s}$. 
In the interested { irreversible phase}, 
$\delta r_{\rm s}>0$, $\tau_{\rm R} \sim 0$ and $\tau_{\rm L} > t_{\rm max}$ with $t_{\rm max}$  the maximum simulation time, 
suggesting that the system reaches a stationary state (see  Fig.~\ref{fig:cAQS}b).
In addition, the MSD in the irreversible phase shows  typical diffusive behavior $\langle \Delta r^2(t) \rangle \sim t$  (see Fig.~\ref{fig:cAQS}c).
In contrast, in the { reversible phase},   
$\delta r_{\rm s}=0$, which means that the system is ``absorbed'' into an invariant state. In practice, one defines $\delta r_\infty = \delta r (t_{\rm max})$ 
and distinguish between the reversible and irreversible phases by comparing the value of $\delta r_\infty$ to a threshold  $\delta r_{\rm th}$. In this study, we set $t_{\rm max} = 4000$ and $\delta r_{\rm th} =0.1$, giving the boundary between irreversible (yellow) and reversible  (blue) regions in Fig.~\ref{fig:cAQS}a.


The above results imply that the configurational space is effectively explored by the dynamical trajectory of the system in the irreversible phase, encouraging us to consider a statistical ensemble generated by shear dynamics. 
{In this study},
we fix $\gamma_{\rm max} = {0.7}$ {unless otherwise specified}, and vary $\varphi$ systematically in the window of $[0.833, 0.849]$. At each $\varphi$, in total $t_{\rm max} \times N_{\rm s}$ independent configurations are generated to construct the ensemble. In the SI, we present additional results obtained for $\gamma_{\rm max} = {1}$. 
{All configurations are collected at $\gamma =0$,  which have weak anisotropy characterized by a non-zero average macroscopic friction coefficient (the ratio between stress and pressure), $\langle \sigma/P \rangle \approx 0.08$.}




Figure~\ref{fig:cAQS}d  shows the evolution of the coordination number  $Z(t)$  in the irreversible phase, obtained from a typical simulation near $\varphi_{\rm J}$. 
Rattlers (particles  with fewer than $d+1$ contacts) are removed in the computation of $Z$.
At first glance, one sees the 
 coexistence of jammed ($Z \approx 2d = 4$) and unjammed states ($Z=0$), 
  similar
 to the coexistence of two ferromagnetic states (positive $m$ and negative $m$) in the  time evolution of the magnetization $m(t)$ in an Ising model near a first-order phase transition~\cite{binder1987finite}.
 However, a more careful examination reveals four distinct states,  which we discuss {in the next section.}

\begin{figure*}[!htbp]
  \centering
  \includegraphics[width=0.6\linewidth]{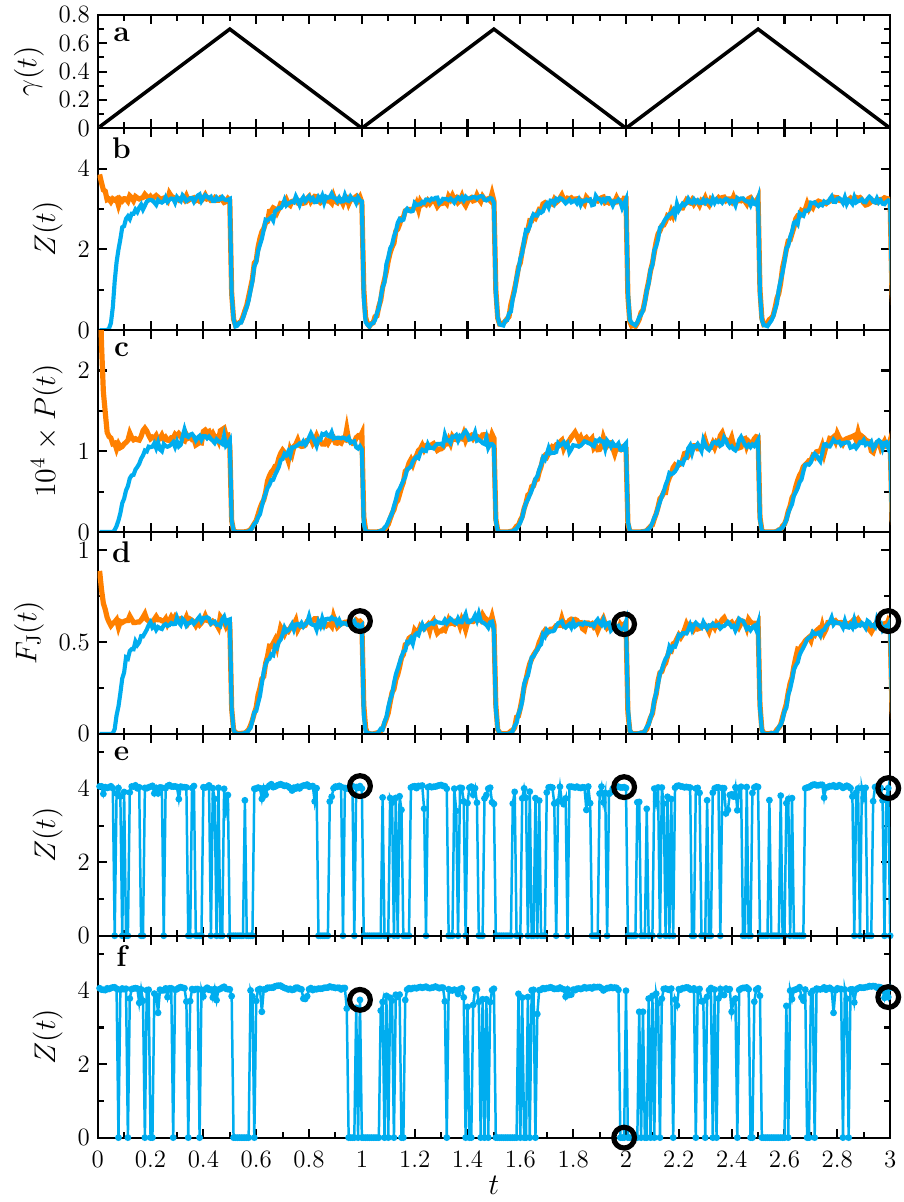}
  \caption{
  {\bf First three shear cycles.} The (a) strain $\gamma$, (b) sample-averaged coordination number $Z$, (c) sample-averaged pressure $P$, and (d) jamming fraction $F_{\rm J}$  are plotted as functions of $t$ during the first three shear cycles ($\varphi = 0.8425$, $N_{\rm s}=768$ samples).
  Orange and blue curves are obtained from rapidly quenched ($\varphi_{\rm j}=\varphi_{\rm J} \approx 0.843$) and mechanically trained ($\varphi_{\rm j}\approx0.848$) initial configurations respectively. 
 The jamming fraction $F_{\rm J}$ used in the scaling analysis is obtained at $\gamma =0$ (black cycles in (d)).
In (e) and (f), plotted is the $Z(t)$  of two typical samples initially generated by rapid quenching.
The system status at $\gamma = 0$ is highlighted by black cycles, which fluctuate among 
 jammed ($Z \geq 4$), fragile ($3<Z<4$) and unjammed ($Z=0$) states.
  \color{black}
  }
  \label{fig:z_time}
\end{figure*}

It is well known  now that the jamming density $\varphi_{\rm j}$ is not unique and protocol-dependent~\cite{parisi2010mean, chaudhuri2010jamming, ozawa2012jamming,  kumar2016memory,  kawasaki2020shear, das2020unified,   babu2021dilatancy, jin2021jamming} (In this study,  $\varphi_{\rm J}$ denotes the J-point density that is the minimum jamming density, and 
$\varphi_{\rm j}$ denotes the protocol-dependent jamming density):
In the compression protocol, $\varphi_{\rm j}$ depends on the compression rate or the density of the  initial equilibrium configuration~\cite{chaudhuri2010jamming, ozawa2012jamming}; in the cyclic athermal  quasistatic compression (CAQC) protocol, $\varphi_{\rm j}$ depends on the volumetric strain amplitude (the maximum density $\varphi_{\rm max}$ to which the system is compressed) and the number $N_{\rm c}$ of cycles~\cite{kumar2016memory,  kawasaki2020shear};  in the CAQS protocol, $\varphi_{\rm j}$ depends on the shear strain amplitude $\gamma_{\rm max}$ and $N_{\rm c}$~\cite{das2020unified,   babu2021dilatancy}. 
The continuous range of possible $\varphi_{\rm j}$ is called a { jamming-line} (J-line), and the minimum jamming density on the J-line is the J-point density $\varphi_{\rm J}$ realized by rapid quenching~\cite{babu2021dilatancy}. 
The protocol-dependence of $\varphi_{\rm j}$
in the current model
 has been investigated in Ref.~\cite{kawasaki2020shear} using the CAQC protocol, and the maximum jamming density obtained there is  $\varphi_{\rm j} \approx 0.8465$ after one over-compression cycle, compared to $\varphi_{\rm J} \approx 0.842-0.843$. In order to examine the effects of non-universal $\varphi_{\rm j}$, in this study we prepare two types of samples: rapidly quenched samples with $\varphi_{\rm j} =  \varphi_{\rm J} \approx 0.843$, and mechanically trained samples generated by the CAQC protocol with 
 $\varphi_{\rm j} = {0.8479(4)} > \varphi_{\rm J}$.

 The protocol-dependence of $\varphi_{\rm j}$ is called a { memory effect} in Ref.~\cite{kumar2016memory}. Below we demonstrate that the cyclic shear simulation employed in this study erases the memory of the initial state: given the initial $\varphi_{\rm j} \approx 0.848 > \varphi_{\rm J}$, the jamming density reduces to $\varphi_{\rm J} \approx 0.843$ after one complete shear cycle. 
Figure~\ref{fig:z_time}b-d shows the evolution of the sample-averaged coordination number $Z$, the sample-averaged pressure $P$, and the jamming fraction  $F_{\rm J}$ of all considered samples at the given $t$. The differences between the curves of rapidly quenched  ($\varphi_{\rm j}=\varphi_{\rm J} \approx 0.843$) and mechanically trained systems ($\varphi_{\rm j} \approx 0.848$) only appear in the first cycle. 
Thus the effect of the variable $\varphi_{\rm j}$ is eliminated during the construction of the ensemble by  cyclic shear. 
Previous studies have shown that, independent of the initial jamming density $\varphi_{\rm j}$, if the system
is sheared by large strain deformations beyond the yield strain $\gamma_{\rm Y}$, 
it always evolves into a critical steady state that has a generic jamming density at
$\varphi_{\rm J}$, and such a process is  accompanied with significant shear dilatancy or hardening  effects~\cite{babu2021dilatancy, pan2023shear, xing2024origin}. 
According to Fig.~\ref{fig:z_time}b-d, the system has indeed reached the steady state before the strain reversal, as the strain amplitude $\gamma_{\rm max} = 0.7$ is  larger than the yield strain $\gamma_{\rm Y} \sim  0.1$.

Next we illustrate that the statistical properties of the ensemble  are determined by fluctuation effects. 
The overall behavior of the { sample-averaged} $Z(t)$ and $P(t)$ is very similar to that reported in previous cyclic shear simulations~\cite{kumar2016memory, luding2021jamming}: one observes periodic patterns with the sharp onset of unjamming upon strain reversal at integer or half-integer $t$. However, no regular patterns can be identified in { single-sample} curves, which look more like stochastic processes (see Fig.~\ref{fig:z_time}e-f). 
Similar behavior has been observed in previous simple shear simulations~\cite{heussinger2009jamming} and experiments~\cite{behringer2008granular}, showing that the  coexistence of jammed and unjammed states during shear is a generic property of the system, at a fixed $\varphi$ near the jamming transition.
It is useful to consider two types of fluctuations. (i) Fluctuations between different cycles in the simulation of a given sample.
The configurations at consecutive strain steps $t$ and $t+\delta t$ are related by both affine and non-affine deformations. Due to non-affine deformations, the two finite-size configurations can have different structures and jamming densities $\varphi_{\rm J}^N$. If the preset constant density $\varphi$ in the CAQS simulation is close to the average $\varphi_{\rm J}$, then the two consecutive configurations can be either jammed ($\varphi_{\rm J}^N<\varphi$) or unjammed ($\varphi_{\rm J}^N>\varphi$). This fluctuation effect is the origin of stochastic-like behavior in Fig.~\ref{fig:z_time}e-f. (ii) Sample-to-sample fluctuations at a given $t$.
For similar reasons, at a given $t$, two individual samples can be either jammed or unjammed due to sample-to-sample fluctuations (see marked points at the integer $t$ in Fig.~\ref{fig:z_time}e-f). The jamming fraction $F_{\rm J}$ due to such sample-to-sample fluctuations is plotted in Fig.~\ref{fig:z_time}d, whose scaling behavior is analyzed in detail below. 
Note that the  fluctuations in (i) and (ii) essentially have the same origin: in finite-size systems, the jamming density $\varphi_{\rm J}^N$ has a distribution $p(\varphi_{\rm J}^N)$ due to the existence of abundant amorphous states.

As shown in Fig.~\ref{fig:z_time}d, upon on the strain reversal, $F_{\rm J}$ rapidly decays to zero, which means that the configuration jammed along a given shear  direction is unstable to reverse shear deformations~\cite{kumar2016memory, boschan2019jamming, seto2019shear}. This reversal-unjamming effect occurs within a strain interval $\delta \gamma \lesssim 0.2$, and after this interval $F_{\rm J}$ reaches a  stationary plateau. Because the maximum strain $\gamma_{\rm max} > 0.2$ in our protocol, the configurations collected in the ensemble are always in the stationary regime.

\color{black}

{\bf Four states: unjammed, jammed, partially crystallized and fragile.}

\begin{figure*}[!htbp]
  \centering
  \includegraphics[width=\linewidth]{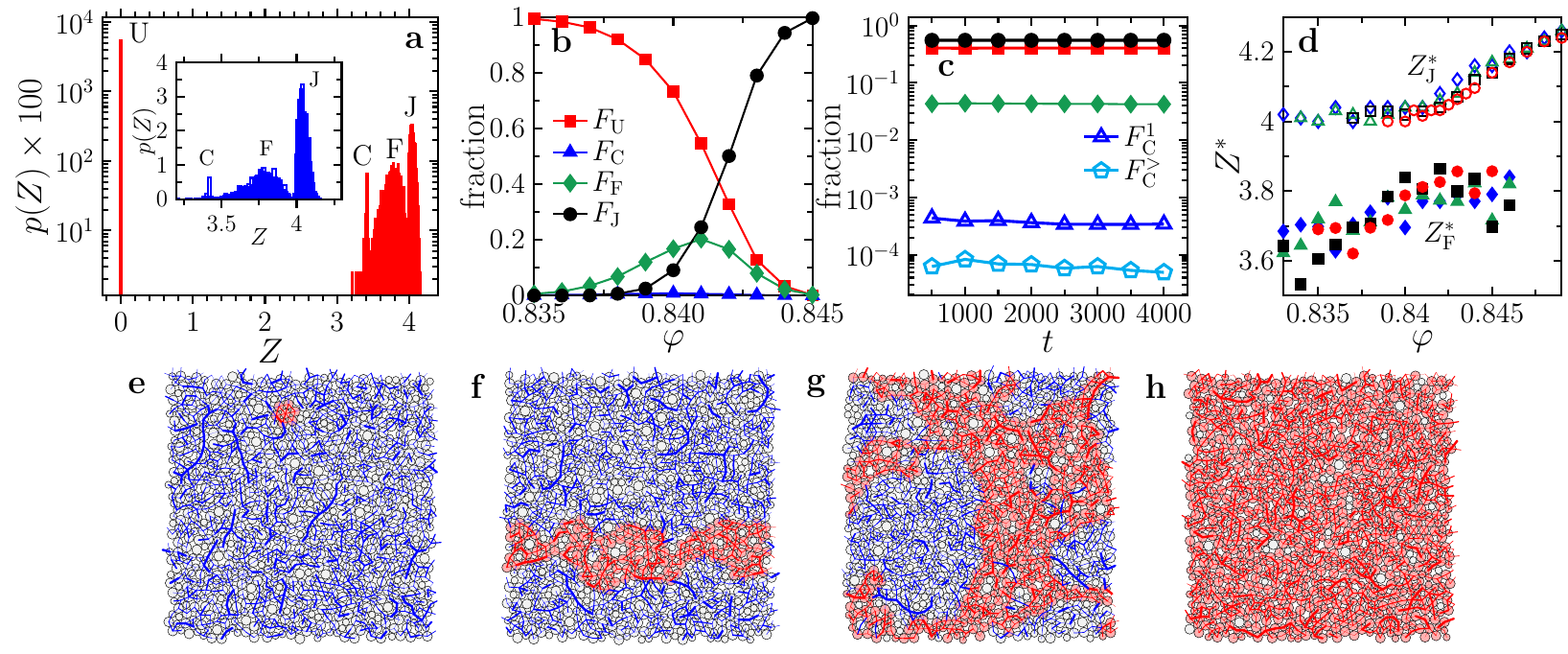}
  \caption{{\bf Four states.} (a) Semi-log plot of the probability distribution $p(Z)$. The  four peaks of unjammed (U), partially crystallized (C), fragile (F) and jammed (J) states, are  indicated.  The inset shows a linear plot of $p(Z)$ without the unjammed peak. 
   Fractions are plotted as a function of $\varphi$ in (b), and of $t$ in (c).
In (c), each data point 
is obtained for a small window  $[t-\delta t, t+\delta t]$ with $\delta t= {250}$.  
   (d) Peak coordination numbers $Z_{\rm J}^*$ of jammed states (upper branch, open symbols) and $Z_{\rm 
   F}^*$ of fragile states (lower branch, filled symbols) as functions of $\varphi$, for $N=256$ (diamonds), 512 (triangles), 1000 (squares), and 2000 (circles).
   In (e-h), we show typical configurations of (e) partially crystallized, (f) fragile percolating in one direction, (g) fragile percolating in both directions, and (h) jammed states.
    Contact forces are represented by bonds, whose width  is proportional to the magnitude of  force.
    The red and blue disks are  
    non-rattler and rattler particles respectively. 
  We set  $\varphi = 0.841$ for (a) and (c), $N=256$ for (b) and (c), and $N=1000$ for ({a}, e-h).
  }
  \label{fig:4states}
\end{figure*}



In Fig.~\ref{fig:4states}a, we plot the probability distribution $p(Z)$ of the states in the considered ensemble, 
at a fixed $\varphi = 0.841$ in the irreversible phase. The following four states, represented by peaks in $p(Z)$, can be identified (see Figs.~\ref{fig:4states}e-h). 


(i) {\it Unjammed states.} The left-most peak is a delta-function $p_{\rm U}(Z) \sim \delta(Z)$, corresponding to unjammed states. 
All unjammed states have strictly zero contacts, $Z=0$, once rattlers are removed.

(ii) {\it Jammed states.} The right-most  peak $p_{\rm J}(Z)$ at $Z\geq4$ corresponds to jammed states.  Their average coordination numbers satisfy minimally the isostatic condition, $Z \geq 2d = 4$.

(iii) {\it Partially crystallized states.}  The delta-peak at $p_{\rm C}(Z) \sim \delta(Z-24/7)$ represents the states with a single unit cell of the hexagonal crystal (see Fig.~\ref{fig:4states}e). Here $24/7 \approx 3.4286$ is the average number of contacts of the seven particles forming the unit cell. 
Occasionally,  states with two or three unit crystal cells can be also found but they are  rare.

(iv) {\it Fragile states.} We define the states in the broad peak $3<Z<4$, excluding the crystalline peak $p_{\rm C}(Z)$, as the fragile states. 

The fractions, $F_{\rm U}$, $F_{\rm J}$, $F_{\rm C}$ and $F_{\rm F}$, 
of the above four states (i-iv), are obtained by integrating corresponding peaks in $p(Z)$. In Fig.~\ref{fig:4states}b, the fractions are plotted as functions of $\varphi$, showing that $F_{\rm J}(\varphi)$ increases from zero to one across  $\varphi_{\rm J}$.
This behavior is quantitatively  similar to $F_{\rm J}(\varphi)$ obtained in previous rapid quench simulations, where finite-size analyses have been carried out to precisely determine the 
 asymptotic jamming transition density  $\varphi_{\rm J}^\infty$ in the thermodynamic limit $N \to \infty$~\cite{o2003jamming, vaagberg2011finite}. We will perform such finite-size analyses later, after discussing the nature of fragile states.

In Fig.~\ref{fig:4states}c, we report various fractions as  functions of $t$ at $\varphi = {0.841}$. The fractions are independent of $t$, confirming that the system is stationary. We further divide $F_{\rm C}(t)$ into two parts,  $F_{\rm C}(t) = F_{\rm C}^1(t) + F_{\rm C}^>(t)$, where $F_{\rm C}^1(t)$ and $F_{\rm C}^>(t)$  are respectively the fractions of partially crystallized states with one and more than one  crystal unit cells.
 Both $F_{\rm C}^1(t)$  and $F_{\rm C}^>(t)$ are  independent of  $t$, indicating that the growth of seed crystals is not observed. 
Because $F_{\rm C}$ 
is generally several orders of magnitude lower than the fractions of other types,  partially crystallized states will be ignored in the following analyses. 

In Fig.~\ref{fig:4states}d, we plot the maximum coordination numbers, $Z_{\rm J}^*$ and $Z_{\rm F}^*$, of jammed and fragile peaks. Around $\varphi_{\rm J}$, there is a small gap $\delta Z_{\rm gap} \approx 0.2$ between $Z_{\rm J}^*$ (upper branch) and $Z_{\rm F}^*$ (lower branch). The results are very similar to those obtained by quasistatic uniform shear in Ref.~\cite{heussinger2009jamming} (see Fig.~4 therein). In Ref.~\cite{heussinger2009jamming}, it is suggested that $\delta Z_{\rm gap}$ vanishes in the thermodynamic limit, and thus the jamming transition is continuous. However, as we show below, the fragile states are generated due to incomplete energy minimization and are mechanically unstable. Once such fragile states are excluded, the lower branch only contains unjammed states with $Z_{\rm U}^*=0$, which is separated by 
a large gap $\delta Z_{\rm gap} \approx 4$ from the upper branch $Z_{\rm J}^* \geq 4$.\\

{\bf Instability of fragile states.}

Previously, the fragile states were obtained by { uniform shear} at non-zero strains $\gamma > 0$ below $\varphi_{\rm J}$ in experiments~\cite{bi2011jamming} and simulations~\cite{kumar2016memory}. 
It was proposed that fragile and jammed states differ in the percolation of the strong force network: in fragile states the percolation occurs anisotropically in one direction only, while in jammed states  it occurs isotropically in both directions.  
{In this study, all states are collected  during  { cyclic shear} at $\gamma=0$, and}
the fragile states with $3<Z<4$ can have percolated force networks of non-rattlers in one or two directions (see Fig.~\ref{fig:4states}f and g). Thus in our case, anisotropy 
 cannot effectively distinguish  fragile from jammed states.
 {It should be noted that, due to strong fluctuation effects, the fragile states can be generated either before or after jamming during the shear procedure (see Fig.~\ref{fig:z_time}e-f).}

\begin{figure}[!htbp]
  \centering
  \includegraphics[width=\linewidth]{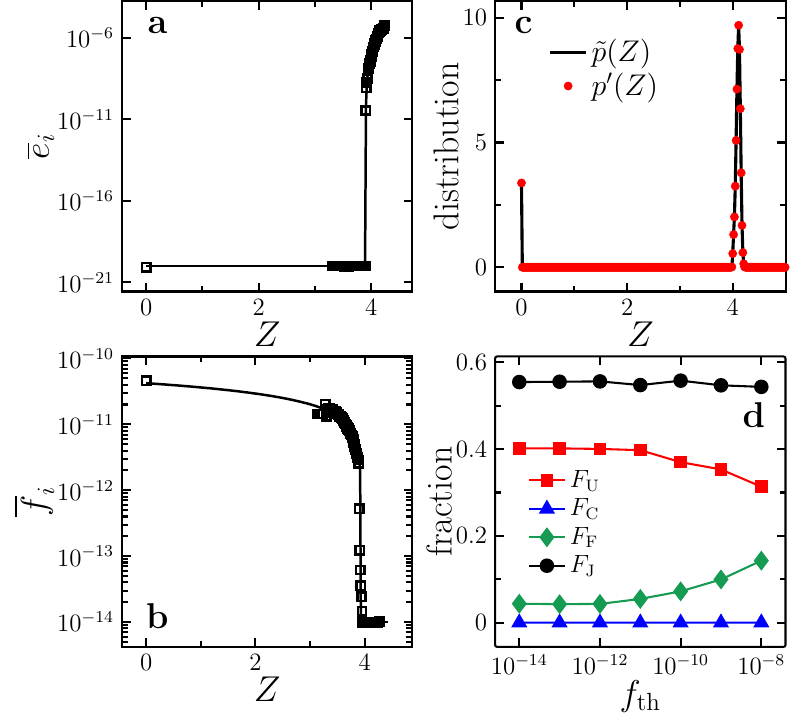}  
  \caption{{\bf Stability tests.} Average single-particle (a) potential energy $\overline{e}_i$ and (b) net force $\overline{f}_i$ as  functions of $Z$. Lines in (a) and (b) are guides to the eye. 
  (c) Comparison between $p'(Z)$ obtained after the compression-decompression perturbation and $\tilde{p}(Z)$ with fragile states counted as unjammed states, for $N=1000$ at $\varphi = 0.844$.
  (d) Fractions of four states as functions of $f_{\rm th}$.
  Data are obtained for  $N=256$ systems {at $\varphi = 0.841$}, except for (c).}
  \label{fig:fragile}
\end{figure}

We demonstrate that the essential difference between fragile and jammed states is their mechanical stability.
In fragile states,  the potential energy  per particle is  negligibly low (see {\it Stopping criterion for energy minimization} in Methods), $\overline{e}_i < e_{\rm th} =  10^{-20}$ (see Fig.~\ref{fig:fragile}a), but the fraction of non-rattler particles is non-zero. These non-rattlers experience forces, which can form a transient, percolated network.  Such networks are highly heterogeneous (see Fig.~\ref{fig:4states}f and g), compared to those in jammed states (see Fig.~\ref{fig:4states}h). More importantly, the force networks in fragile states are unstable, revealed by the non-zero net force per particle, $\overline{f}_i > f_{\rm th} =  10^{-14}$  (see Fig.~\ref{fig:fragile}b). Thus fragile states are not strictly equilibrated; they turn into unjammed states by sufficiently  long relaxation (accurate energy minimization) or mechanical perturbations.  

To demonstrate the instability of fragile states,  two tests are carried out. 
First, we perform a compression-decompression perturbation, $\varphi \to \varphi + \delta \varphi \to \varphi$, where  $\delta \varphi  = 10^{-8}$, with the energy  minimized after each step. All fragile states ($3<Z<4$) become unjammed ($Z=0$) after this perturbation, while jammed states remain. In Fig.~\ref{fig:fragile}c, we plot the distribution $p'(Z)$ after the compression-decompression perturbation. Independently, without any perturbation we simply regard all fragile states ($3<Z<4$) as  unjammed ($Z=0$) and recalculate the distribution $\tilde{p}(Z)$. Figure~\ref{fig:fragile}c  shows that the two distributions $p'(Z)$ and $\tilde{p}(Z)$ perfectly coincide. 

In the second test, we repeat CAQS simulations by systematically varying the threshold $f_{\rm th}$ in the criterion of the energy minimization algorithm.  The fraction $F_{\rm F}$ of fragile states grows and $F_{\rm U}$ of unjammed states decays with increasing $f_{\rm th}$ (Fig.~\ref{fig:fragile}d),  suggesting that many unjammed states become fragile under a looser force-balance condition $\overline{f}_i < f_{\rm th}$ with a larger  $f_{\rm th}$.  In contrast, the constant   $F_{\rm J}$ shows that the definition of  jammed states is insensitive to the algorithm parameter. 
{The  $p(Z)$ data with different $f_{\rm th}$ confirm this property: the fragile peak depends on $f_{\rm th}$ while the jammed peak is independent of $f_{\rm th}$ (see SI Fig.~S1a).}
Note that in granular experiments~\cite{bi2011jamming, zhao2019shear}, the friction may play the role of $f_{\rm th}$, i.e., the net inter-particle force could be balanced by the frictional force between particles and the supporting plate. According to this assumption, the probability of observing fragile states  in experiments would depend on the particle-plate friction that can be changed by replacing the materials. It provides a protocol to examine our scenario of fragile states in future experimental studies.  

Because fragile {(and partially crystallized)} states {with $3<Z<4$} are unstable, from now on we count them as unjammed states. More specifically, we replace the original distribution $p(Z)$ with the modified distribution $\tilde{p}(Z)$. For simplicity, we omit the tilde and denote $\tilde{p}(Z)$ by  $p(Z)$ below. \\

\begin{figure*}[!htbp]
  \centering  
  \includegraphics[width=0.9\linewidth]{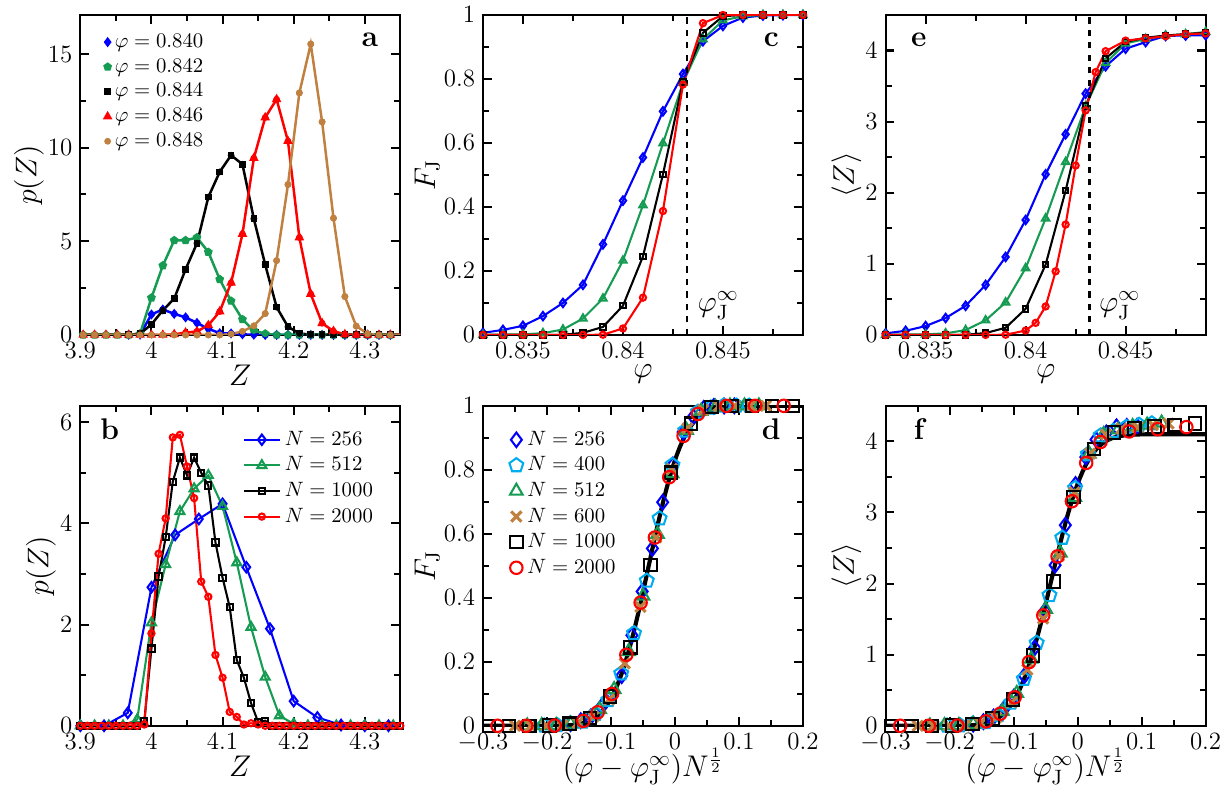} 
  \caption{{\bf Distribution of the coordination number.} 
  (a) Probability distribution $p(Z)$ at a few different $\varphi$ for $N=1000$. (b) $p(Z)$ at  $\varphi = 0.843$ for a few different $N$. 
  For better visualization, we do not show the unjammed delta-peak at $Z=0$ {(see SI Fig.~S1b for full distributions)}. (c) The fraction of jammed states $F_{\rm J}$ as a function of $\varphi$ for a few different $N$. The intersection of curves gives $\varphi_{\rm J}^\infty = {0.8432(2)}$.
  (d) The data points of $F_{\rm J}$ with different $N$ collapse as a function of $(\varphi - \varphi_{\rm J}^\infty)N^{1/2}$. The solid line represents fitting to Eq.~(\ref{eq:FJ}), with two fitting parameters $u = {0.041(1)}$ and $\sigma_\varphi ={ 0.043(1)}$.  The average coordination number $\langle Z \rangle$ is plotted as a function of  $\varphi$ in (e) and of $(\varphi - \varphi_{\rm J}^\infty)N^{1/2}$ in (f). The solid line in (f) represents $\langle Z \rangle = F_{\rm J} Z_{\rm J}^*$ using $u = 0.041$, $\sigma_\varphi = 0.043$ and $Z_{\rm J}^* = 4.1$. Symbols in (b-f) have the same meanings.
  }
  \label{fig:pZ}
\end{figure*}

{\bf Scaling analysis  near the jamming transition.}



Firstly, we  outline the general strategy in the following scaling analyses. In particular,  the effects of disorder on a first-order phase transition shall be specified.
Near a first-order phase transition, the probability distribution $p(m)$ of the order parameter $m$ has two well-separated peaks corresponding to two phases~\cite{binder1987finite, binder1987theory}, $p(m) = p_1(m) + p_2(m)$.
The fractions of the two phases are respectively, $F_1 = \int p_1(m) dm$ and  $F_2 = \int p_2(m) dm$, satisfying $F_1 + F_2 = 1$.  
In the thermodynamic limit $N \to \infty$, $F_1(x)$  and $F_2(x)$ are sharp step functions of the control parameter $x$ (for example, the temperature $T$), which jump discontinuously at the transition point $x_{\rm c}$. In finite-size systems, $F_1(x, N)$ and $F_2(x, N)$ turn into 
 smooth functions due to fluctuations, and usually a finite-size scaling analysis becomes  necessary to determine the nature of the transition. 
Near $x_{\rm c}$, the fraction
$F_1(x, N)$ (or $F_2(x, N)$) follows a scaling form, $F_1(x, N) = \mathcal{F}_1 [ (x - x_{\rm c}) N^\lambda]$, where the value of $\lambda$ depends on  whether disorder presents. 
It is also illustrative to study the behavior of 
the { connected susceptibility} $\chi_{\rm con} = d\langle m \rangle/dx$ and the { disconnected susceptibility}
$\chi_{\rm dis} = N \left( \langle m^2 \rangle - \langle m \rangle^2 \right)$. The scalings are unambiguously distinguishable in the following two types of first-order phase transitions. 

(I) Standard first-order transitions without disorder. 
In this case, $F_1(x, N)$ has the scaling form $F_1(x, N) = \mathcal{F}_1 [ (x - x_{\rm c}) N ]$.
The fluctuation-dissipation theorem ensures that the susceptibility can be equivalently computed  from the response or from the fluctuations,
i.e., $\chi_{\rm dis} = \chi_{\rm con}$.
Examples in this category include the
first-order transition in the
$q$-state Potts model with $q>4$ where the control parameter is the temperature $T$, and that in the Ising model driven by an external magnetic field $h$ at temperatures below the critical temperature.

(II) First-order transitions in disordered systems driven by an external field, where sample-to-sample fluctuations due to disorder are significantly stronger than  thermal fluctuations and the system can be effectively considered as athermal ($T=0$).
The scaling form of $F_1(x, N)$ is  $F_1(x, N) = \mathcal{F}_1 [ (x - x_{\rm c}) N^{1/2} ]$. The exponent $\lambda=1/2$ originates from the finite-size scaling of the standard deviation of the transition point $\delta x_{\rm c} \sim N^{-1/2}$ due to the presence of disorder.
The two susceptibilities are
related by, $\chi_{\rm dis} \sim \chi_{\rm con}^2 \sim N$~\cite{kierlik2002adsorption}, where the average $\langle \cdots \rangle$ in the definition of susceptibilities should be taken over disorder realizations.
A paradigm  transition in this type  is the non-equilibrium first-order transition in the athermal random field Ising model (RFIM) driven by an external field $h$~\cite{nattermann1998theory,  rossi2023effective}. 
Other examples  include 
the brittle yielding  in amorphous solids under strain $\gamma$ deformations~\cite{ozawa2018random, rossi2023effective}, and the melting of ultrastable hard-sphere glasses under decompression~\cite{zhang2022machine}.

For the interested jamming transition, the control parameter is $\varphi$,  {by tuning which continuously the jamming transition occurs}, and the order parameter is $Z$, {which characterises the jammed ($Z>4$) and unjammed phases ($Z=0$)}.
{They are}
analogous to the external field $h$ and the magnetization $m$ in the athermal driven RFIM in type (II). 
The transition is between unjammed and jammed phases, represented by two separated peaks in the  distribution $p(Z)$.
The  fractions of unjammed and jammed states are $F_1(\varphi, N) = F_{\rm U}(\varphi, N)$ and $F_2(\varphi, N) = F_{\rm J}(\varphi, N)$ respectively. The disconnected susceptibility is defined as, $\chi_{\rm dis} \equiv N  \sigma_Z^2= N \langle  (Z - \langle Z \rangle)^2 \rangle$, where $ \sigma_Z^2$ is the variance of  $p(Z)$  and $\langle \ldots \rangle$ is the  average over all states.  The connected susceptibility is defined as $\chi_{\rm con} =  d\langle Z \rangle/d\varphi$. Below we analyze in detail the simulation data of  $F_{\rm J}(\varphi, N)$,
$\chi_{\rm dis}(\varphi, N)$ and $\chi_{\rm con}(\varphi, N)$,  demonstrating that they satisfy the scalings in (II). Note that the disorder may have different origins in type (II) transitions. In RFIM, the disorder corresponds to the quenched random local fields in the definition of the Hamiltonian, while for the jamming transition, the disorder is due to the amorphous configuration of particle arrangements.

\color{black}

In Figs~\ref{fig:pZ}a and b, we plot  $p(Z)$ for a few different $\varphi$ and $N$, showing that $p(Z)$ is always double-peaked across the jamming transition {(see SI Fig.~S1b)},
which is typical  behavior of a first-order rather than a second-order transition. 
To analyze the scaling behavior of $p(Z)$, we consider a general form of first-order phase transitions~\cite{binder1987finite, binder1987theory},
\beq
p(Z)   = (1-F_{\rm J})  \delta(Z)  + F_{\rm J}  \, p_{\rm J}\left[ (Z- Z_{\rm J}^*)N^\eta \right],
\label{eq:pZ}
\eeq
where  $\delta(Z)$ and $p_{\rm J}(Z)$ correspond to the unjammed and jammed  peaks respectively. 
The asymptotic jamming density $\varphi_{\rm J}^{\rm \infty} = {0.8432(2)}$ for $N \to \infty$ is determined by the intersection of $F_{\rm J}(\varphi)$ curves with different $N$ (see Fig.~\ref{fig:pZ}c).
{This value is 
consistent with  the J-point density $\varphi_{\rm J} \approx 0.842-0.843$ given by previous studies~\cite{o2003jamming, vaagberg2011finite}, which is unambiguously below the 
maximum protocol-dependent jamming density $\varphi_{\rm j} \approx 0.8465$ reported in Ref.~\cite{kawasaki2020shear} and $\varphi_{\rm j} \approx 0.848$ obtained by CAQC in this study.}
The difference between $\varphi_{\rm J}^{\rm \infty} = {0.8432(2)}$ by $\gamma_{\rm max} =0.7$ and $\varphi_{\rm J}^{\rm \infty} = {0.8435(1)}$ by $\gamma_{\rm max} =1$ (see SI Fig.~{S2b and {S3-4}}) is too small to conclude any systematic dependence on $\gamma_{\rm max}$. 
In Ref.~\cite{das2020unified}, an unjamming pocket is reported in the phase diagram  for $\gamma_{\rm max} < 0.17$, while for larger $\gamma_{\rm max}$, $\varphi_{\rm J}$ is independent of $\gamma_{\rm max}$, consistent with our observations. 
These results also suggest that the jamming density obtained by cyclic shear is the lowest jamming density on the J-line of all possible jammed states~\cite{parisi2010mean, chaudhuri2010jamming, babu2021dilatancy}.  {We emphasize that  $\gamma_{\rm max} = 0.7$ in this study is significantly larger than the yield strain $\gamma_{\rm Y} \approx 0.1$ (see SI Fig.~S5a). 
For a 
small strain amplitude, $\gamma_{\rm max} = 0.05$, we observe a logarithmic growth of $\varphi$ as a function of $t$ under a constant pressure condition~\cite{knight1995density, kumar2016memory}, suggesting that the jamming density can be increased by cyclic shear with a small $\gamma_{\rm max}$~\cite{das2020unified, babu2021dilatancy}; in contrast, for $\gamma_{\rm max} = 0.7$, $\varphi$ remains as a constant (see SI Fig.~S5b)}

As shown in Fig.~\ref{fig:4states}d, $Z_{\rm J}^*$ is independent of $N$ near $\varphi_{\rm J}$, and weakly depends on $\varphi$.
In the following scaling analysis, $Z_{\rm J}^*$ is approximated by a constant value $Z_{\rm J}^* {= 4.1(1)}$ at $\varphi_{\rm J}^{\rm \infty} = 0.8432$.
To keep the expressions  concise, we ignore the corrections to the scaling functions from the
$\varphi$-dependence of $Z_{\rm J}^*$. 



We assume the fraction of jammed states $F_{\rm J}$ having  the following scaling form,
$F_{\rm J}(\varphi, N) =  \mathcal{F}_{\rm J} \left [ (\varphi - \varphi_{\rm J}^\infty) N^
\lambda \right]$.
The value of the exponent $\lambda$ is important for understanding the nature of the transition.  
If the system were thermal, 
the fraction would be determined by the Boltzmann distribution, $F_{\rm J} \sim \exp \left(\frac{N \delta f}{k_{\rm B} T}\right)$, where $\delta f$ is the single-particle free energy difference between two phases~\cite{binder1987theory, binder1987finite}. Because the free energy is non-singular around a first-order phase transition, it can be expanded, giving $\delta f \sim (T - T_{\rm c})$ to the lowest order. Thus, if the jamming  transition were a standard first-order phase transition, then  $F_{\rm J}(\varphi, N) = \mathcal{F}_{\rm J}\left [ (\varphi - \varphi_{\rm J}^\infty) N \right]$, i.e., $\lambda = 1$ (note that $\varphi$ {is the control parameter of the jamming transition}).
However, our numerical data can not be collapsed using  $\lambda= 1$; in contrast, they can be perfectly collapsed using  $\lambda = 1/2$ (see Fig.~\ref{fig:pZ}d). 

To understand the origin of $\lambda = 1/2$ scaling, recall that the athermal jamming transition is   not driven by the free energy difference between the two phases. 
Below we explain the scaling by the scenario of a first-order transition with quenched disorder. 
For a finite $N$, due to the  presence of disorder in the packing structure, the jamming density $\varphi_{\rm J}^N$ should fluctuate around the asymptotic value $\varphi_{\rm J}^\infty$. Let us assume a simple Gaussian form of the distribution, 
$
\rho(\varphi_{\rm J}^N) \sim \exp \left[-\frac{(\delta \hat{\varphi}_{\rm J} + u)^2}{2 \sigma_\varphi^2}\right],
$
where $\delta \hat{\varphi}_{\rm J} =(\varphi_{\rm J}^N- \varphi_{\rm J}^\infty)  N^{1/2} $ follows the standard central limit theorem.
Note that $\rho(\varphi_{\rm J}^N)$ has been explicitly measured in the compression protocol~\cite{o2002random,o2003jamming}: the width of  $\rho(\varphi_{\rm J}^N)$  scales as $w\sim N^{-0.55}$ in both 2D and 3D, supporting our assumption.
The above assumption also predicts, $\varphi_{\rm J}^\infty - \varphi_{\rm J}^N \sim N^{-1/2}$, independent of the dimensionality.
The jammed states are defined by $\varphi > \varphi_{\rm J}^N$, and thus
\beq
\begin{split}
F_{\rm J}(\varphi, N) = \int_{0} ^ {\varphi}  \rho(\varphi_{\rm J}^N)  \, d\varphi_{\rm J}^N \approx \frac{1}{2} +  \frac{1}{2} \erf \left[\frac{\delta \hat{\varphi}+ u}{\sqrt{2} \sigma_{\varphi}} \right],
\end{split}
\label{eq:FJ}
\eeq
where $\delta \hat{\varphi} = (\varphi - \varphi_{\rm J}^\infty) N^{1/2}$. Equation~(\ref{eq:FJ}) agrees  well with the data in Fig.~\ref{fig:pZ}d.
{The dependence of $F_{\rm J}(\varphi, N)$ on shear parameters is examined in Fig.~S2. The $F_{\rm J}(\varphi, N)$ data are independent of the maximum strain $\gamma_{\rm max}$, and weakly depend on the strain step $\delta \gamma$. Independent of $\delta \gamma$, the $F_{\rm J}(\varphi, N)$ data of different $N$ collapse when plotted as  a function of  the rescaled variable $(\varphi - \varphi_{\rm J}^\infty)N^{1/2}$ (see Figs.~\ref{fig:pZ}d and~S2c for $\delta \gamma = 10^{-2}$ and $10^{-1}$ respectively), showing robustness of the scaling function Eq.~(\ref{eq:FJ}).}

 

\begin{figure*}[!htbp]
  \centering  
  \includegraphics[width=0.9\linewidth]{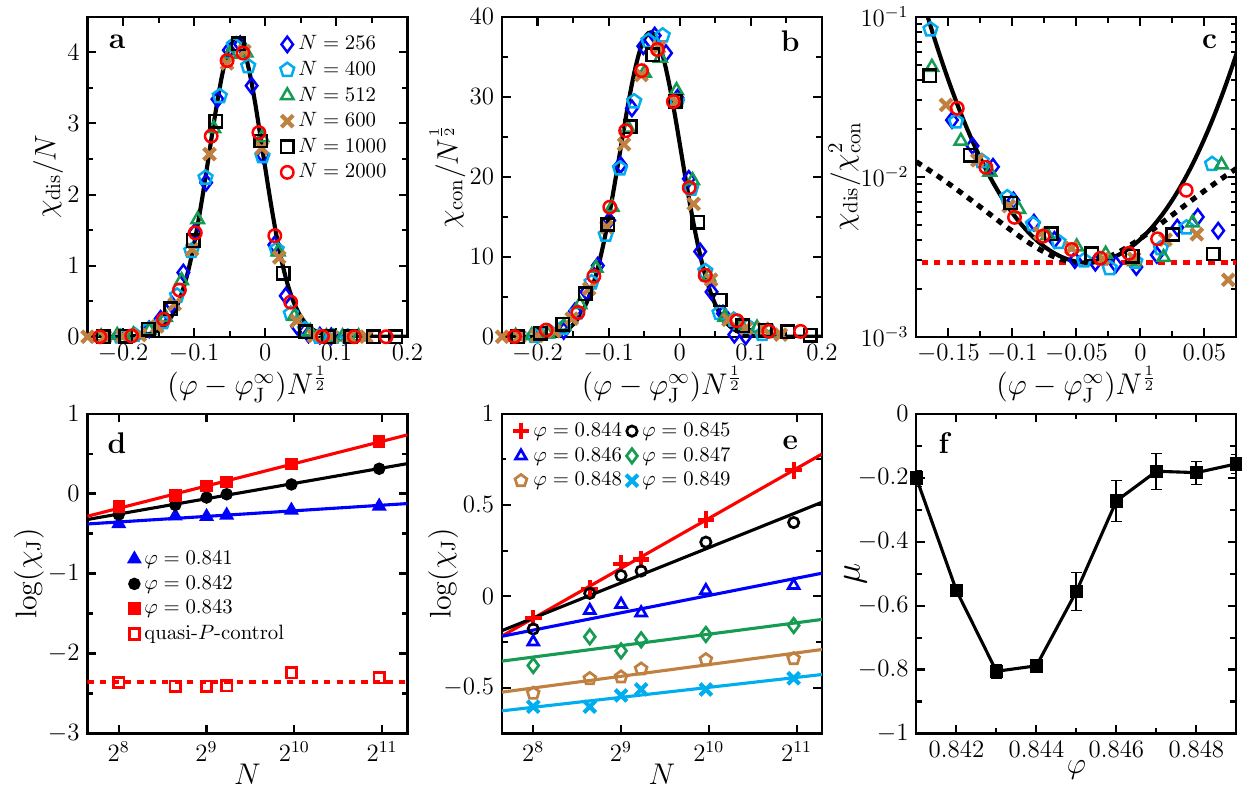} 
  \caption{{\bf Susceptibilities.} 
  We plot (a)
  the disconnected susceptibility $\chi_{\rm dis}$ rescaled by $N$, (b) the connected susceptibility $\chi_{\rm con}$ rescaled by $N^{1/2}$, and (c) the ratio between $\chi_{\rm dis}$ and $\chi_{\rm con}^2$,  as functions of $(\varphi - \varphi_{\rm J}^\infty)N^{1/2}$. The solid black lines in (a-c) are {Eqs.~(\ref{eq:chi_dis}),~(\ref{eq:chi_con}), and the ratio between Eq.~(\ref{eq:chi_dis}) and the square of Eq.~(\ref{eq:chi_con}), respectively.
  The dashed black line in (c) is the second-order expansion Eq.~(\ref{eq:relation}).} The dashed red line in (c) is the constant term in Eq.~(\ref{eq:relation}), i.e., $\chi_{\rm dis}/\chi_{\rm con}^2 = \pi \sigma_\varphi^2/2$. To draw these lines, we use 
  {$u = 0.041$}
  and $\sigma_\varphi=0.043$ obtained from the fitting in Fig.~\ref{fig:pZ}d, and $Z_{\rm J}^*=4.1$ determined at $\varphi_{\rm J}^\infty = 0.8432$ in Fig.~\ref{fig:4states}d. No fitting is performed here.
Symbols in (a-c) have the same meanings as indicated in the legend of (a).
  In (d) and (e), the susceptibilities of jammed states $\chi_{\rm J}$ are plotted as functions of $N$ for a few different $\varphi$, and are fitted to Eq.~(\ref{eq:chiJ}). 
   The open squares in (d) are $\chi_{\rm J}$ data obtained in
   a small pressure window {$0.0009 < P < 0.0011$}, corresponding to 
   {$\Delta Z \approx 0.15$ ($\varphi \approx 0.843 - 0.844$)}.
  The fitting parameter $\mu$ is plotted in (f) as a function of $\varphi$, where the error bar represents the fitting error.
  }
  \label{fig:chi}
\end{figure*}

Once the scaling behavior of $F_{\rm J}(\varphi, N)$ is obtained, it is easy to derive scaling forms of susceptibilities.
From Eq.~(\ref{eq:pZ}) {and the definition of $\chi_{\rm dis}$}, 
we obtain,
\beq
\frac{\chi_{\rm dis} (\varphi, N)}{N} \approx (Z_{\rm J}^*)^2 [1 - F_{\rm J}(\varphi, N)]F_{\rm J}(\varphi, N),
\label{eq:chi_dis}
\eeq
where $F_{\rm J}(\varphi, N)$ is given by Eq.~(\ref{eq:FJ}). Equation~(\ref{eq:chi_dis}) suggests a scaling form $ \chi_{\rm dis}(\varphi, N)/N= \mathcal{X}_{\rm dis} \left[ (\varphi -  \varphi_{\rm J}^\infty)  N^{1/2} \right]$, in agreement with the data  in Fig.~\ref{fig:chi}a.



From Eq.~(\ref{eq:pZ}), one finds that $\langle Z \rangle \approx F_{\rm J} Z_{\rm J}^*$ and thus $\langle Z \rangle$ has the same scaling form as $F_{
\rm J}$, i.e., $\langle Z \rangle (\varphi, N) = \mathcal{Z}[(\varphi -  \varphi_{\rm J}^\infty)  N^{1/2}]$, consistent with the data in 
Fig.~\ref{fig:pZ}e and~\ref{fig:pZ}f.
Using  $\chi_{\rm con} = Z_{\rm J}^* \frac{d F_{\rm J}} {d\varphi}$ and Eq.~(\ref{eq:FJ}), we obtain,
\beq
\frac{\chi_{\rm con}(\varphi, N)}{N^{1/2}} \approx \frac{Z_{\rm J}^*}{\sigma_\varphi\sqrt{2\pi}} \exp \left[ - \left( \frac{\delta \hat{\varphi}+ u}{\sqrt{2} \sigma_{\varphi}} \right)^2 \right],
\label{eq:chi_con}
\eeq 
which is verified by the  simulation data in Fig.~\ref{fig:chi}b.


Now we can look at the relationship between 
$\chi_{\rm dis}$ and  $\chi_{\rm con}$. Expanding Eqs.~(\ref{eq:chi_dis}) and ~(\ref{eq:chi_con}) around the maxima, where $x \equiv \frac{\delta \hat{\varphi}+ u}{\sqrt{2} \sigma_{\varphi}}=0$, we obtain, up to the quadratic order,
\beq
\frac{\chi_{\rm dis}}{\chi_{\rm con}^2} \approx
\frac{\pi \sigma^2_{\varphi}}{2} \left[1+ \left(2-\frac{4}{\pi}\right) \left( \frac{\delta \hat{\varphi}+ u}{\sqrt{2} \sigma_{\varphi}} \right)^2 \right].
\label{eq:relation}
\eeq 
To the lowest order, Eq.~(\ref{eq:relation}) gives a scaling,
$
\chi_{\rm dis} \sim \chi_{\rm con}^2,
$
which is a key signature of the presence of quenched disorder. 
Comparing to the simulation data, Eq.~(\ref{eq:relation}) works well around the extreme point (see Fig.~\ref{fig:chi}c).
{The agreement can be improved  using higher-order expansions.}

It seems that the 
non-Gaussian effect in the distribution $\rho(\varphi_{\rm J}^N)$,  which is neglected in the present analysis, is amplified in the data of $\chi_{\rm dis}/\chi_{\rm con}^2$, resulting in slight asymmetry. 
In SI Figs.~{S3-4 and S6-7}, we show that the scaling functions Eqs.~(\ref{eq:FJ})-(\ref{eq:relation}) work in 3D, and for a different max strain $\gamma_{\max} = 1$ in 2D.
{In 3D, we obtain $\varphi_{\rm J}^\infty = {0.6479(2)}$, consistent with the previously reported J-point density $\varphi_{\rm J} \approx 0.648$ of this  model~\cite{das2020unified}, which is the minimum jamming density on the J-line. For the same 3D model, the protocol-dependent jamming density is obtained up to $\varphi_{\rm j} = 0.6616$ by the thermal annealing method~\cite{chaudhuri2010jamming}, and up to $\varphi_{\rm j} = 0.661$ by the CAQS mechanical training method~\cite{das2020unified}.}


Next let us discuss the finite-size effects of the jammed peak $p_{\rm J} (Z)$ in Eq.~(\ref{eq:pZ}). 
For standard first-order phase transitions, 
$\eta = 1/2$~\cite{binder1987theory, binder1987finite}.   
Thus in that scenario the susceptibility of the jamming peak would scale as, 
$\chi_{\rm J} \equiv N  \sigma_{\rm J}^2= N \langle  (Z - Z_{\rm J}^*)^2 \rangle_{\rm J} {\sim {\rm constant}}$,
where $ \sigma_{\rm J}^2$ is the variance of  $p_{\rm J}(Z)$  and $\langle \ldots \rangle_{\rm J} $ represents the  average restricted to the jammed states only. 
However, our  $\chi_{\rm J}$ data disagree with this scaling (see Figs.~
\ref{fig:chi}d-f). 
At different $\varphi$, $\chi_{\rm J}$ can be  fitted to  a  pow-law form {$\chi_{\rm J}  \sim \ell^{-\mu}$~\cite{hexner2018two, hexner2019can}, or}  
\beq
\chi_{\rm J}  \sim N^{-\mu/d},
\label{eq:chiJ}
\eeq
where {$N \sim \ell^d$. The exponent} $\mu=d(2 \eta-1)$ is a fitting parameter. 
At large or small $\varphi$ away from $\varphi_{\rm J}^{\rm \infty}$ (see Fig.~
\ref{fig:chi}f), the exponent is close to zero ($\mu \approx -0.2$), suggesting that the local coordination numbers $Z_i$ are uniformly distributed.  
The deviation from the uncorrelated behavior ($\mu = 0$) is  significant around $\varphi_{\rm J}^{\rm \infty}$, where $\mu \approx -0.8$ reaches the minimum.


The result of $\mu$  shall be interpreted with care. In general, (i) $\mu=0$ corresponds to uniformity of $Z_i$, 
(ii) $\mu>0$ to hyperuniformity with a vanishing  $\chi_{\rm J}$ in the thermodynamic limit, and (iii) $\mu<0$ to hyperfluctuations with a diverging  $\chi_{\rm J}$ in the thermodynamic limit that typically appears at the critical point in a second-order phase transition. However, the negative  $\mu$ in Fig.~\ref{fig:chi}f is not due to the criticality of the jamming transition.  
Here it is essential to consider the volume fluctuations. The negative $\mu$ is obtained from a  $\varphi$-controlled setup
in our CAQS simulations. 
If we instead select configurations around a constant  pressure $P$,  then the fluctuations become significantly  smaller, and $\mu \approx 0$ (Fig.~\ref{fig:chi}d). 
This observation is consistent with a previous study~\cite{ikeda2023control}, suggesting that the large fluctuation of $Z$ near $\varphi_{\rm J}^{\rm \infty}$ in the $\varphi$-controlled protocol might be originated from the fluctuation of  $\varphi_{\rm J}^N$. 
In fact, careful measurement gives $\mu \simeq 1$ in a $P$-controlled protocol at jamming, confirming hyperuniformity~\cite{hexner2018two, hexner2019can}. 
{Ref.~\cite{hexner2019can} also compares an ensemble of subsystems cut out from large packings, to an ensemble of whole systems with the same volume under periodic boundary conditions (similar to the $P$-controlled ensemble considered in this study): the  contact fluctuations in the former are larger than those in the latter, and the convergence of these two are expected to occur in extremely large systems that are not attainable in current simulations
-- Due to this pre-asymptotic effect, the hyperuniform exponent $\mu \simeq 1$ is only observable in the first ensemble, while  in the second ensemble, the apparent exponent is close to $\mu \approx 0$, consistent with our $P$-controlled data in Fig.~\ref{fig:chi}d.
In short, although  the hyperuniform exponent $\mu=1$ is not directly observed, our results do not contradict the recently reported hyperuniform behavior of contact distributions at jamming.
}
We emphasize that the finite-size effects of the jammed peak $p_{\rm J} (Z)$ contribute negligibly to the scaling of $F(\varphi, N)$, $\chi_{\rm dis}(\varphi, N)$ and $\chi_{\rm con}(\varphi, N)$. In other words, the first-order nature of the jamming transition is independent of how $Z_i$ is spatially distributed in jammed packings.

{In the scaling analysis presented above,  $Z$ is treated as the essential order parameter.
In principle, the same kind of scaling analysis can be applied to any physical quantity $A$  that varies discontinuously at jamming (e.g., $A$ can be the bulk modulus $B$ or the fraction of non-rattlers $f_{\rm NR}$). The distribution $p(A)$ should have the same form as Eq.~(\ref{eq:pZ}), $p(A) = (1-F_{\rm J})\delta(A) + F_{\rm J} p_{\rm J}(A)$. 
 With fragile states removed, generally $A=0$ below jamming because no contacts are formed  (e.g., $B=0$ and $f_{\rm NR}=0$), and thus the unjammed peak is always a delta function. 
The fraction of jammed states $F_{\rm J}$ only depends on the constructed ensemble, which determines how the states are sampled, and thus its scaling form $F_{\rm J}(\varphi, N)=\mathcal{F}_{\rm J}[(\varphi - \varphi_{\rm J}^\infty) N^{1/2}]$ is independent of $A$. The jammed peak $p_{\rm J}(A)$ does depend on $A$, but as shown above, $p_{\rm J}(A)$ is irrelevant to the interested scaling behavior of $F_{\rm J}, \chi_{\rm dis}$ and $\chi_{\rm con}$. Thus, our conclusion drawn from the scaling analysis should be robust and generic, independent of which parameter $A$ is chosen for the scaling analysis, as long as $A$ can reflect the discontinuous nature of jamming. 
}
{If one instead chooses a continuous variable, such as the pressure $P$, then the above scaling analysis would not work. }
\\

{\Large \bf Discussion}

In this study, we {investigate}
the nature of  {the} jamming transition through  an ensemble approach analogous to the statistical mechanics in equilibrium systems~\cite{baule2018edwards, bi2015statistical}. Within such a framework, {jamming is demonstrated to be a first-order transition with quenched disorder,
in the quasistatic deformation limit where  all fragile states are excluded.
The} order of the jamming transition is independent of the complex properties of jammed packings, including isostaticity, marginality and hyperuniformity (see (i-iii) in Introduction). 

{In previous rapid quench simulations, it is observed that the width $w$ of the  jamming density distribution $P(\varphi_{\rm J}^N)$  scales as $w \sim N^{-\Omega}$ with $\Omega = 0.55 \pm 0.03$ in both 2D and 3D~\cite{o2003jamming}. This finite-size scaling depends on the total number of particles, $N$, rather than on the system length $\ell$, and the exponent  $\Omega \approx 1/2$ is independent of the dimensionality within the numerical accuracy. An explanation, as suggested previously, is that jamming is a second-order transition with an upper critical dimension $d_{\rm u} = 2$~\cite{goodrich2012finite}. Here we propose an alternative interpretation: jamming is a first-order transition with quenched disorder, and consequently the finite-size scaling is a function of $(\varphi - \varphi_{\rm J}^\infty)N^{1/2}$, independent of the dimensionality.}
{We expect that the same finite-size scaling applies to the steady state configurations generated by uniform simple shear.}

It is of particular interest to reconcile the first-order transition established here under quasi-static shearing and the second-order transition observed  in previous finite-rate rheology (see (iv) in Introduction). 
{Conventionally, first- and second-order transitions coexist in gas-liquid systems, but not in liquid-crystalline solid systems. Because the liquid-crystalline solid transition is accompanied with spontaneous symmetry breaking, it cannot be a continuous type. However, there is no apparent  spontaneous symmetry breaking during liquid-amorphous solid transitions (such as the jamming transition). Thus there is no fundamental reason to exclude the possibility of the coexistence of discontinuous  and continuous transitions between a liquid and an amorphous solid. This study opens an avenue to unify first-order (this study) and second-order~\cite{olsson2007critical} jamming transitions. 

{In this study, we ignore thermal fluctuations. An interesting}
 direction of the future work is to study the competition between the sample-to-sample fluctuations due to disorder and the fluctuations associated to the { granular temperature}~\cite{poschel2001granular, luding2021jamming}, and examine the ``thermal effects'' on the nature of the first-order jamming transition.    
Similar  studies on the thermal vestiges of the zero-temperature physics have been recently carried out in the driven RFIM at finite temperatures~\cite{yao2023thermal}.
It would be also interesting to examine if  friction can alter the order of the jamming transition~\cite{song2008phase, ramaswamy2023universal}. 
}\\

{\Large \bf Methods}

{\it Models.}
We study models of frictionless, bidisperse particles in two  and three  dimensions.
The number ratio between large and small particles is 1:1, and the diameter ratio is 1.4:1.
The potential energy between two particles is:
\beq
U(r_{ij})=\frac{\epsilon}{2}\left(1-\frac{r_{ij}}{\sigma_{ij}} \right)^2\Theta \left(1-\frac{r_{ij}}{\sigma_{ij}} \right), 
\label{eq:U_r}
\eeq
where $\epsilon=1$ is the energy unit, $r_{ij}$  the distance between particles $i$ and $j$, $\sigma_{ij}=\frac{\sigma_i+\sigma_j}{2}$ the mean diameter, and  $\Theta(x)$ the Heaviside step function. 
We set unit particle mass, and  the {mean diameter}
as the unit length. \\

{\it Shear protocol.}
Particles are randomly distributed at an initial volume fraction $\phi_0=0.02$. The system is  then rapidly quench compressed to the target $\varphi$. 
The CAQS is performed under the Lees-Edwards boundary conditions~{\cite{lees1972computer}}, with a fixed $\varphi$.
{The strain $\gamma$ is varied stepwise, rather than continuously, between 0 and $\gamma_{\rm max}$.}
We use a  strain step $\delta \gamma = 0.1$ to generate the phase diagram in Fig.~\ref{fig:cAQS}a, and $\delta \gamma= 0.01$ for other results. 
At each step, particle positions are shifted according to $x_i\to x_i+\delta \gamma y_i$, and then the system's energy is minimized using the FIRE algorithm~\cite{bitzek2006structural}.  
During a cycle, the strain is varied as $\gamma = 0 \to \gamma_{\rm max} 
\to 0$. 
We set $\gamma_{\rm max} = 0.7$ in the main text and  $\gamma_{\rm max} = 1$ in the SI for the 2D model, and $\gamma_{\rm max} = 0.5$ for the 3D model (SI).
The number of cycles is represented by $t$
with unit oscillation period ($t=1$).  
In 2D, the maximum number of cycles is $t_{\rm max}=250$ for $0.839 \leq \varphi \leq {0.849}$ and $t_{\rm max}=4000$ for ${0.833} \leq \varphi < 0.839$, while in 3D, {$t_{\rm max} = 144$ for $2000$, and $t_{\rm max} = 50$ for $N=512$ and $1000$}.
We generate 
$N_{\rm s} = 4000-12000$ independent samples 
at each $\varphi$.
At each $\varphi$, in total $t_{\rm max} \times N_{\rm s}$ configurations are collected 
to compute  statistical quantities.\\

{\it Stopping criterion for energy minimization.}
We terminate the energy minimization when the potential energy  per particle $\overline{e}_i = \frac{1}{N}\sum_{i=1}^{N} e_i$ falls below a threshold $e_{\rm th}$ or the average single-particle net force $\overline{f}_i = \frac{1}{N}\sum_{i=1}^{N} f_i$ falls below $f_{\rm th}$, whichever is satisfied earlier. 
We set $e_{\rm th} = 10^{-20}$ and $f_{\rm th} =  10^{-14}$ unless otherwise specified.\\

{\bf Data availability}

Source data are provided with this paper.\\

{\bf Code availability}

The simulation program is available
at https://github.com/dengpan-cn/Library-of-Athermal-Quasi-static-Shear-Simulation-of-Granular-Matter.\\

\vspace{1cm}

{\bf Acknowledgments}

We warmly thank Hajime Yoshino, {Stephen Teitel, Bulbul Chakraborty, Olivier Dauchot, Gilles Tarjus, Yujie Wang} for inspiring discussions. 
We acknowledge financial support from NSFC (Grants 12161141007, 11935002,
11974361 and 12047503), from 
Chinese Academy of Sciences (Grants ZDBS-LY-7017 and KGFZD-145-22-13), 
and from Wenzhou Institute (Grant WIUCASQD2023009). In this work access was granted to the High-Performance Computing Cluster of Institute of Theoretical Physics - the Chinese Academy of Sciences.\\

{\bf Author Contributions Statement}

All authors contributed equally to this work. \\

{\bf Competing Interests Statement}

The authors declare no competing interests.\\


\clearpage
\onecolumngrid
\centering {\LARGE \bf Supplementary Information}
\setcounter{figure}{0}
\setcounter{equation}{0}
\setcounter{table}{0}
\setcounter{section}{0}
\renewcommand\thefigure{S\arabic{figure}}
\renewcommand\theequation{S\arabic{equation}}
\renewcommand\thesection{S\arabic{section}}
\renewcommand\thetable{S\arabic{table}}

\begin{figure*}[!htbp]
	\centering  
	\includegraphics[width=0.8\linewidth]{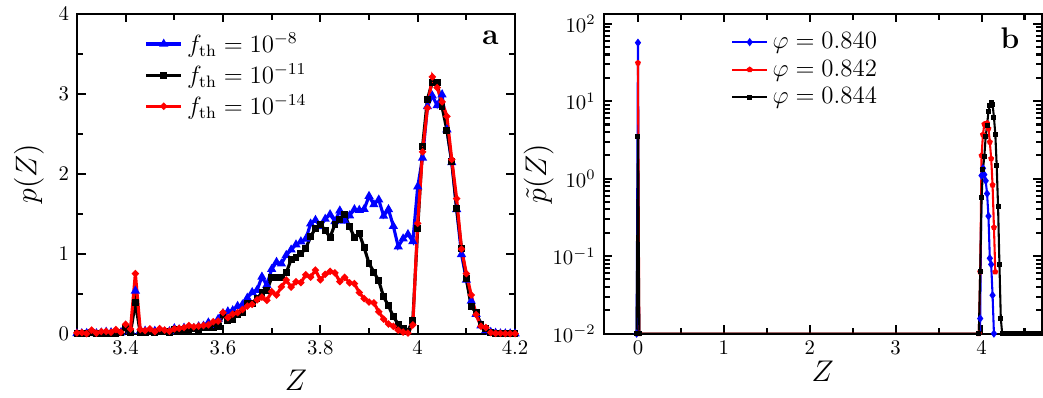} 
	\caption{{\bf Additional data on the distribution $p(Z)$ in two dimensions.} 
		(a) $p(Z)$ for a few different $f_{\rm th}$ ($\varphi = 0.841$ and $N=1000$). 
		(b)
		Full probability distribution $\tilde{p}(Z)$ at a few different $\varphi$ near $\varphi_{\rm J}^\infty = 0.8432$ ($N=1000$ and $f_{\rm th} =10^{-14}$). 
	}
	\label{fig:PZ_double_peak}
\end{figure*}

\begin{figure*}[!htbp]
	\centering  
	\includegraphics[width=\linewidth]{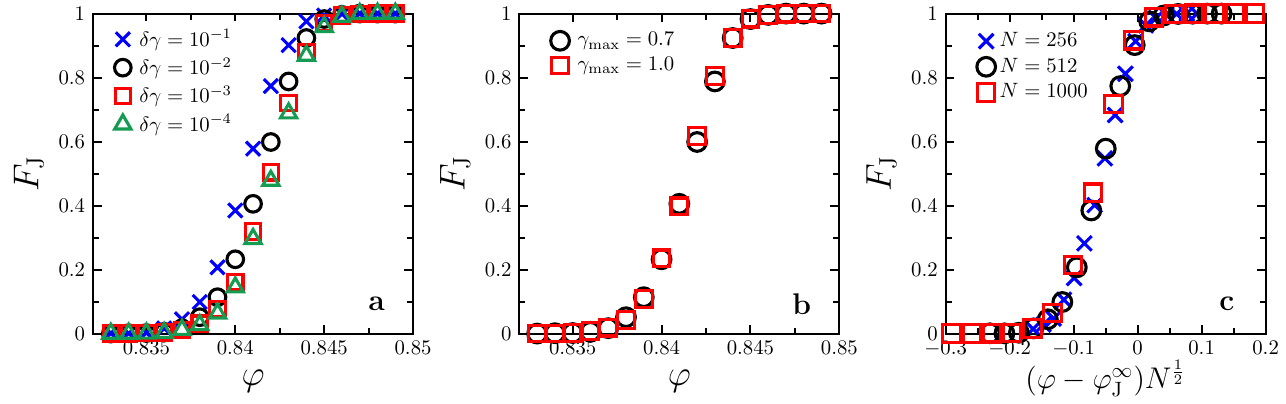} 
	\caption{{\bf Additional data on the jamming fraction $F_{\rm J}$ in two dimensions.}
		(a) $F_{\rm J}$ as a function of $\varphi$, for different strain steps $\delta \gamma$, with fixed $\gamma_{\rm max} = 0.7$ and $N=512$.
		(b) $F_{\rm J}$ as a function of $\varphi$, for different maximum strains $\gamma_{\rm max}$, with fixed $\delta \gamma = 10^{-2}$ and $N=512$.
		(c) Data collapse of $F_{\rm J}(\varphi, N)$ as a function of rescaled variable $(\varphi - \varphi_{\rm J}^\infty)N^{1/2}$, 
		with $\varphi_{\rm J}^\infty = 0.8432$, $\delta \gamma = 10^{-1}$ and $\gamma_{\rm max} = 0.7$.
	}
	\label{fig:FJ_parameter_dependence}
\end{figure*}

\begin{figure*}[!htbp]
	\centering  
	\includegraphics[width=0.7\linewidth]{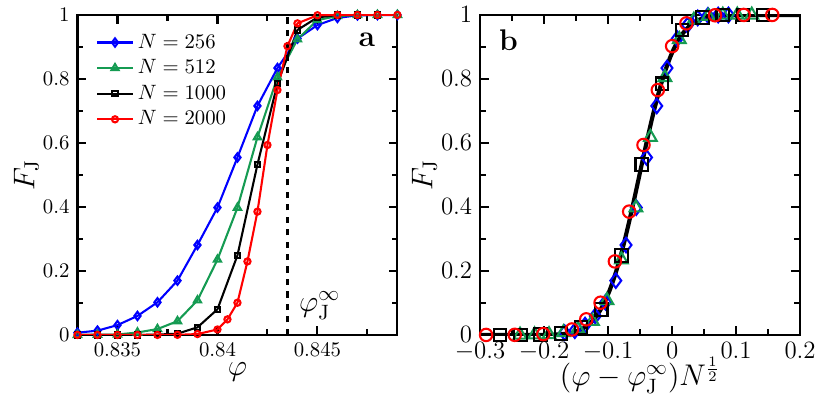} 
	\caption{{\bf 
			{Jamming fraction} 
			in two dimensions   for $\gamma_{\rm max}=1$.} 
		(a) Fraction of jammed states $F_{\rm J}$ as a function of $\varphi$ for a few different $N$. The intersection of curves gives $\varphi_{\rm J}^\infty = {0.8435(1)}$.
		(b) The data points of $F_{\rm J}$ with different $N$ collapse as a function of $(\varphi - \varphi_{\rm J}^\infty)N^{1/2}$. The solid line represents the fitting according to Eq.~(4),
		with two fitting parameters $u = {0.051(1)}$ and $\sigma_\varphi = {0.043(1)}$. 
	}
	\label{fig:pZ_g1}
\end{figure*}

\begin{figure*}[!htbp]
	\centering  
	\includegraphics[width=0.9\linewidth]{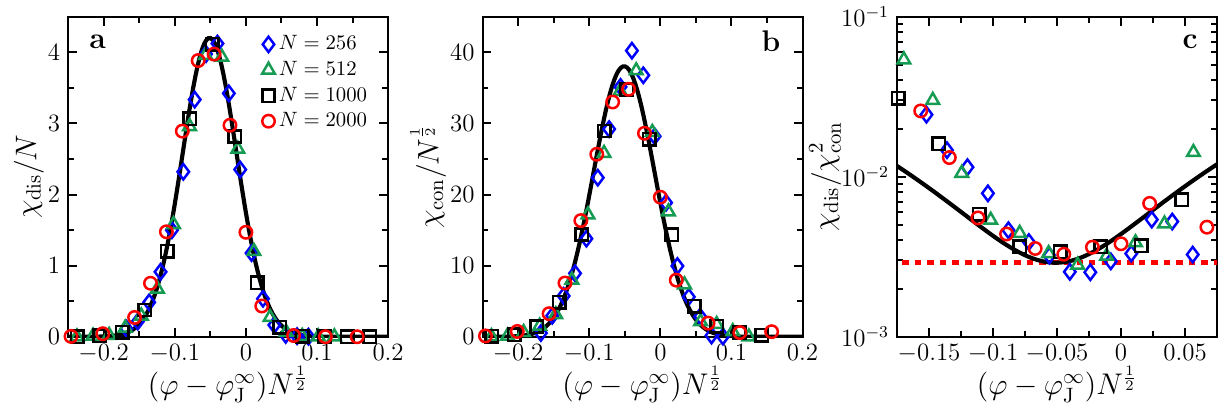} 
	\caption{{\bf Susceptibilities in two dimensions  for $\gamma_{\rm max}=1$.} 
		(a)
		The disconnected susceptibility $\chi_{\rm dis}$ rescaled by $N$, (b) the connected susceptibility $\chi_{\rm con}$ rescaled by $N^{1/2}$, and (c) the ratio between $\chi_{\rm dis}$ and $\chi_{\rm con}^2$ are plotted as functions of $(\varphi - \varphi_{\rm J}^\infty)N^{1/2}$. The solid black lines in (a-c) are Eqs.~(5),~(6) and (7) 
		respectively. The dashed red line in (c) is the constant term in Eq.~(7), 
		i.e., $\chi_{\rm dis}/\chi_{\rm con}^2 = \pi \sigma_\varphi^2/2$. To draw these lines, we use $u = 0.051$ and $\sigma_\varphi=0.043$ obtained from the fitting in Fig.~\ref{fig:pZ_g1}b.
		Symbols in (a-c) have the same meanings as indicated in the legend of (a). 
	}
	\label{fig:chi_g1}
\end{figure*}

\begin{figure*}[!htbp]
	\centering  
	\includegraphics[width=0.75\linewidth]{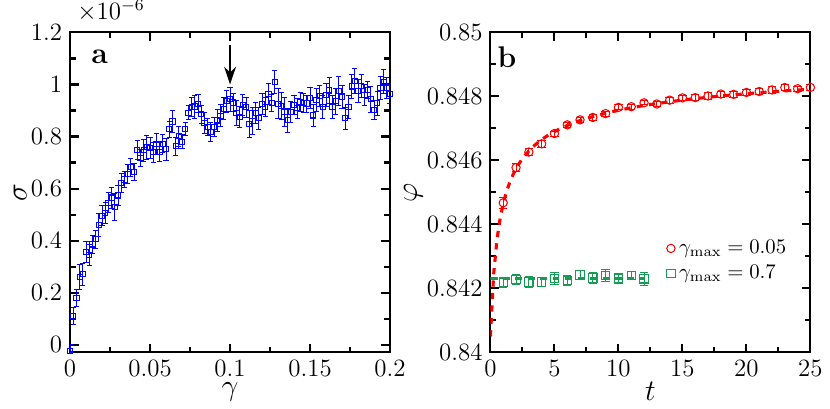} 
	\caption{{{\bf Constant-pressure shear simulations in two dimensions.} 
			(a) Stress $\sigma$ as a function of the strain $\gamma$ in simple athermal quasistatic shear, under a constant pressure $P=10^{-5}$. The yield strain is $\gamma_{\rm Y} \approx 0.1$ (arrow),
			beyond which the curve becomes nearly a plateau.
			(b) $\varphi$ as a function of the number of 
			cycles $t$ in cyclic athermal quasistatic shear (in one cycle $\gamma$ varies as $0 \to \gamma_{\rm max} \to -\gamma_{\rm max} \to 0$), with a fixed $P=10^{-5}$. The data of $\gamma_{\rm max} = 0.05$ are fitted to a logarithmic form (dashed line), $\varphi(t) = \varphi_{\rm f} - \frac{\varphi_{\rm f} - \varphi_0}{1 + B\log(1 + t/\tau)}$, as proposed in Ref.~[74].
			Here $\varphi_0 = 0.8405$ is the packing fraction of un-strained initial systems;   $\varphi_{\rm f} = 0.8516(5)$, $B = 0.61(15)$ and $\tau = 0.60(18)$ are fitting parameters.}
		The error bar represents the standard error of the mean.
	}
	\label{fig:constant_P}
\end{figure*}

\begin{figure*}[!htbp]
	\centering  
	\includegraphics[width=0.7\linewidth]{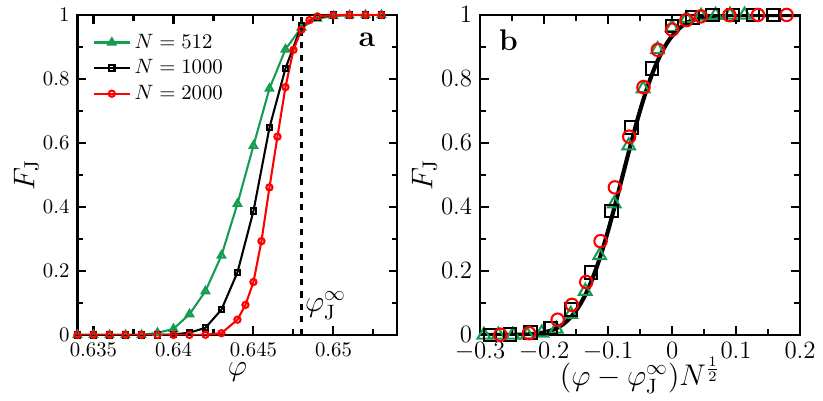} 
	\caption{{\bf 
			{Jamming fraction}
			in three dimensions.} 
		(a) Fraction of jammed states $F_{\rm J}$ as a function of $\varphi$ for a few different $N$. The intersection of curves gives $\varphi_{\rm J}^\infty = {0.6479(2)}$.
		(b) The data points of $F_{\rm J}$ with different $N$ collapse as a function of $(\varphi - \varphi_{\rm j}^\infty)N^{1/2}$. The solid line represents the fitting according to Eq.~(4),
		with two fitting parameters $u = {0.077(2)}$ and $\sigma_\varphi = {0.051(1)}$. 
	}
	\label{fig:pZ3D}
\end{figure*}

\begin{figure*}[!htbp]
	\centering  
	\includegraphics[width=0.9\linewidth]{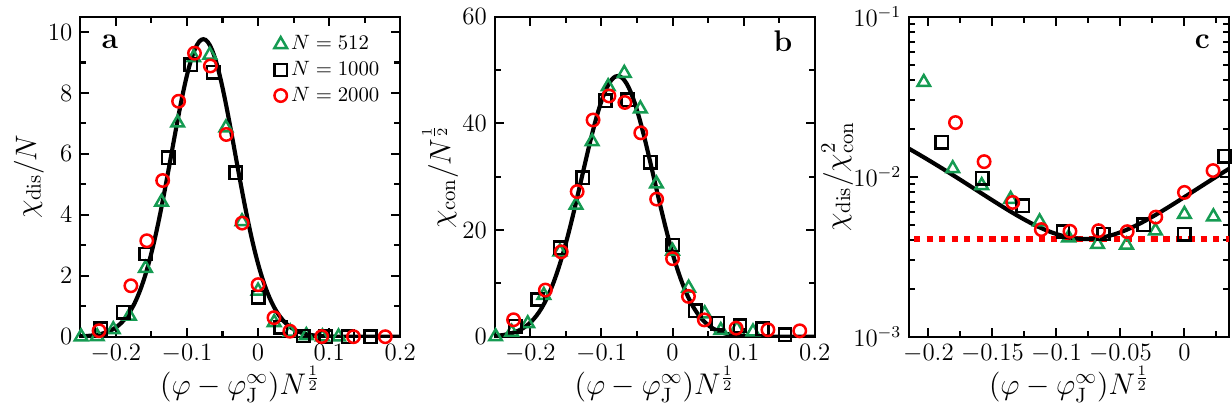} 
	\caption{{\bf Susceptibilities in three dimensions.} 
		(a)
		The disconnected susceptibility $\chi_{\rm dis}$ rescaled by $N$, (b) the connected susceptibility $\chi_{\rm con}$ rescaled by $N^{1/2}$, and (c) the ratio between $\chi_{\rm dis}$ and $\chi_{\rm con}^2$ are plotted as functions of $(\varphi - \varphi_{\rm J}^\infty)N^{1/2}$. The solid black lines in (a-c) are Eqs.~(5), (6) and (7) 
		respectively. The dashed red line in (c) is the constant term in Eq.~(7), 
		i.e., $\chi_{\rm dis}/\chi_{\rm con}^2 = \pi \sigma_\varphi^2/2$. To draw these lines, we use $u = 0.077$ and $\sigma_\varphi=0.051$ obtained from the fitting in Fig.~\ref{fig:pZ3D}b. 
		Symbols in (a-c) have the same meanings as indicated in the legend of (a).
	}
	\label{fig:chi3D}
\end{figure*}


\end{document}